\documentclass[sigconf,screen,authorversion,nonacm]{acmart}
\AtBeginDocument{%
  }

\copyrightyear{2025}
\acmYear{2025}
\setcopyright{cc}
\setcctype{by-nc-nd}
\acmConference[ICCSIT 2025]{The 18th International Conference on Computer Science and Information Technology}{October 27--29, 2025}{Paris, France}
\acmBooktitle{The 18th International Conference on Computer Science and Information Technology (ICCSIT 2025), October 27--29, 2025, Paris, France}
\acmPrice{}
\acmDOI{10.1145/3783862.3783873}
\acmISBN{979-8-4007-1858-8/2025/10}

\begin{document}

\title{How Warm Glow Alters the Usability of Technology}

\author{Antonios Saravanos}
\email{saravanos@nyu.edu}
\orcid{0000-0002-6745-810X}
\affiliation{%
  \institution{New York University}
  \streetaddress{7 East 12th Street, Room 625B}
  \city{New York}
  \state{NY}
  \country{USA}
  \postcode{10003}
}
\renewcommand{\shortauthors}{Saravanos, A.}

\begin{abstract}
As technology increasingly aligns with users’ personal values, traditional models of usability that focus on functionality, specifically effectiveness, efficiency, and satisfaction, may not fully capture how people perceive and evaluate a technology. This study investigates how the warm-glow phenomenon, the positive feeling associated with doing good, shapes perceived usability. We used an experimental approach in which participants evaluated a hypothetical technology under conditions designed to evoke either the intrinsic (i.e., personal fulfillment) or extrinsic (i.e., social recognition) dimensions of warm glow. A multivariate analysis of variance and subsequent follow-up analyses revealed that intrinsic warm glow significantly enhances all dimensions of perceived usability, while extrinsic warm glow selectively influences perceived effectiveness and satisfaction. These findings suggest that perceptions of usability extend beyond functionality and are shaped by how technologies resonate with users’ broader sense of purpose. We conclude by proposing that designers consider incorporating warm-glow effects into technology as a strategic design decision.
\end{abstract}



\maketitle

\section{Introduction and Background}

This paper explores the effect of warm glow on the perceived usability of technology. Warm glow is the feeling realized when one does something good, a concept rooted in economics~\cite{andreoni1989,andreoni1990} and psychology. The literature divides the phenomenon into two dimensions based on the motivation behind the action: extrinsic and intrinsic. Extrinsic warm glow is the feeling experienced when recognized for doing something perceived as “good”, while intrinsic warm glow is the personal satisfaction derived from an altruistic act. It has been shown (e.g., Saravanos et al.~\cite{saravanos2022measuring}) that actions generating warm glow are motivated by a combination of both these aspects. As technologies become increasingly associated with activities (i.e., social innovation) and invoke feelings of warm glow in users, the study of their usability is of both theoretical and practical value.

Prior work has incorporated warm glow into models of technology adoption but has not explicitly isolated its impact on usability as defined by the ISO 9241-11 standard (effectiveness, efficiency, and satisfaction)~\cite{iso9241-11}. For example, Saravanos et al.~\cite{saravanos2022tam3wg} introduced the TAM3 + WG model, which integrates warm glow into Technology Acceptance Model 3. They examined how warm glow affects users’ perceptions of ease of use (which can be mapped to efficiency) and usefulness (which can be mapped to effectiveness), and showed that extrinsic warm glow positively influenced a technology’s perceived usefulness, particularly when it enhanced users’ social image. However, while these findings relate to aspects of perceived usability, they do not address usability as a comprehensive construct or evaluate satisfaction as a distinct dimension. Saravanos et al.~\cite{saravanos2026utaut2wg} also enhanced the Unified Theory of Acceptance and Use of Technology 2 model by incorporating warm glow, which they refer to as the UTAUT2 + WG model. They examined whether warm glow influences users’ perceptions of a technology’s ease of use (i.e., perceived efficiency) and usefulness (i.e., perceived effectiveness). They found that both intrinsic and extrinsic warm glow significantly enhanced users’ usability perceptions. Intrinsic warm glow impacted all three usability dimensions, whereas extrinsic warm glow only influenced performance expectancy and hedonic motivation. Although these two studies did not explicitly examine perceived usability with respect to warm glow, in both experiments, it was found to influence perceived usability. Building on these efforts, here we experimentally investigate the effect of warm glow on perceived usability measured through the criteria of effectiveness, efficiency, and satisfaction.

Usability has long been conceptualized as a tripartite construct encompassing effectiveness, efficiency, and satisfaction, a framework codified in ISO 9241-11~\cite{iso9241-11}. However, the ability of the affective dimension to influence usability has been recognized in the literature. Bargas-Avila and Hornbæk’s~\cite{bargasavila2011} systematic review discusses how user experience research, particularly empirical studies, has evolved over time. Their review focuses on studies that highlight the affective dimension of user experience, specifically examining how emotions, moods, and feelings influence and result from interactions with technology.

As an example of such a study, Kurosu and Kashimura~\cite{kurosu1995} generated 26 screen layout patterns from the same graphical components of an ATM interface and asked 252 participants to rate each layout on both functional usability (apparent ease of use) and aesthetic appeal (visual beauty). They found that apparent usability was significantly more influenced by visual aesthetics than by underlying functional design. While the correlation between beauty and apparent usability was moderate ($r = 0.589$), most traditional usability factors, such as key positioning, operation sequence, and grouping, had little to no correlation with users’ perceptions. For example, optimal glance sequence and operation flow showed correlations close to zero, while the layout’s resemblance to familiar key patterns (such as a telephone keypad) had the highest usability impact ($r = 0.730$). This study underscores that users often conflate visual appeal with functional quality. It highlights the need for designers to focus on not only optimizing usability in a technical sense but also ensuring that interfaces look easy to use to promote positive first impressions and user engagement.

Tractinsky~\cite{tractinsky1997} built on Kurosu and Kashimura’s study by attempting to replicate their findings in a different cultural context, specifically that of Israel. He suggested that their findings may reflect the common assumption that Japanese culture places more value on design aesthetics, arguing that Israeli users would show different results. Tractinsky conducted three controlled experiments. In the first, he directly replicated the Japanese study, using 104 Israeli engineering students as participants. Participants were shown the 26 ATM interface layouts in the Japanese study, which were adapted into Hebrew. They were asked to rate each layout on two separate scales: how easy the system appeared to be to use (apparent usability) and how visually appealing it looked (aesthetics). He found an even stronger correlation between aesthetics and apparent usability ($r = 0.92$) than in the original Japanese study. This contradicted his expectation of a weaker correlation for Israeli users, whom he assumed to be less aesthetically oriented than Japanese users.

To ensure that this strong correlation was not merely the result of methodological flaws, such as participants giving similar answers to consecutive questions, Tractinsky’s second experiment modified the procedure. Here, 81 students were divided into two groups. One group rated aesthetics first, and then usability in a second round; the other group did the reverse. The order of design presentation was randomized in both rounds to minimize response bias. The correlation between aesthetics and apparent usability remained very high ($r = 0.83$), suggesting that the relationship was not a byproduct of how the questions were ordered or presented.

In the third experiment, Tractinsky tested for medium bias, which refers to the possibility that projecting designs in a large classroom setting may influence how participants judged them. This time, 108 participants viewed the interfaces individually on personal computers using custom-built software that randomized both the order of designs and the questions. Again, aesthetics and usability ratings were strongly correlated ($r = 0.92$), reinforcing the validity of the previous findings. This setup also allowed measurement of the time required for participants to make their judgments. Interestingly, they took slightly longer to assess usability than aesthetics, supporting the notion that evaluating ease of use is a more cognitively demanding task.

In all three experiments, participants evaluated the same seven objective design variables (including keypad type, button grouping, and layout configuration) that Kurosu and Kashimura~\cite{kurosu1995} had examined as predictors of inherent usability. While a few of these, such as keypad layout and grouping, were moderately correlated with usability, none exhibited the strength of the aesthetics–usability correlation. Moreover, Israeli users had different preferences from their Japanese counterparts. For example, they associated better usability with telephone-style keypads, while Japanese users favored horizontal layouts. This suggests that even objectively “better” interface components may be culturally interpreted, further complicating efforts to design universally usable systems.

Friedman et al.~\cite{friedman2013} further developed the concept of embedding human values more systematically into the design process through value-sensitive design. It is important to note that value-sensitive design expands the scope of values considered in technology design, moving beyond the traditional emphasis on cooperation (e.g., computer-supported cooperative work) and participation and democracy (e.g., participatory design)~\cite{friedman2013}. Value-sensitive design incorporates a wide range of human values, particularly those with moral importance, such as fairness, justice, human well-being, and virtue, drawing from ethical theories including deontology, consequentialism, and virtue ethics~\cite{friedman2013}. In addition, it considers social conventions, such as protocol standards, and individual preferences, like color choices in user interfaces~\cite{friedman2013}.

While previous studies have explored how warm glow can influence technology acceptance and certain aspects of perceived usability, such as usefulness and ease of use, they have not explicitly examined how both intrinsic and extrinsic warm glow, independently and interactively, affect all three core dimensions of usability defined by ISO 9241-11: effectiveness, efficiency, and satisfaction. Furthermore, prior research has often conflated general positive affect or user experience with usability, rather than isolating the specific contribution of warm-glow motivations. Therefore, this study systematically investigates how distinct types of warm glow shape users’ perceptions of effectiveness, efficiency, and satisfaction, providing a more nuanced understanding of how affective factors and personal values become integral to usability judgments. To investigate this gap, we pose the following three research questions:

\begin{enumerate}
  \item[RQ1:] How does intrinsic warm glow affect perceived usability, specifically in terms of effectiveness, efficiency, and satisfaction?
  \item[RQ2:] How does extrinsic warm glow impact perceived usability across these three dimensions?
  \item[RQ3:] How do intrinsic and extrinsic warm glow interact in shaping perceived usability?
\end{enumerate}

The remainder of the paper is organized as follows. Section~\ref{sec:methods} details the experimental design, participant recruitment, and the instruments used to measure perceived usability and warm glow. It explains how validated scales were selected, and adapted where necessary, and describes the rationale behind the vignette manipulations. Section~\ref{sec:results} presents the results, starting with a multivariate analysis of variance (MANOVA) to examine joint effects, followed by univariate tests and permutation analyses, with bootstrapped confidence intervals to assess robustness. Section~\ref{sec:discussion} offers a theoretical and practical discussion of the findings, summarizes the study’s contributions and limitations, and offers directions for future research.

\section{Materials and Methods}
\label{sec:methods}

In this section, we present the methodological approach employed to examine the influence of warm-glow motivations on users’ perceived usability suggested by Saravanos et al.~\cite{saravanos2022measuring}. First, we describe the study design, in which participants were exposed to a hypothetical technology embedding cues intended to elicit either intrinsic or extrinsic warm-glow responses. Second, we detail the participant recruitment strategy and report key demographic characteristics of the sample. Third, we introduce the study materials and the survey instrument used to measure all constructs of interest. In particular, perceived usability was operationalized along three dimensions, namely perceived efficiency, perceived effectiveness, and perceived satisfaction, as well as the components of warm glow. Finally, we summarize the data collection procedures and outline the statistical analyses conducted to address the research questions.

\subsection{Measures}

To examine the influence of warm-glow motivation on user experience, we assessed two primary constructs: perceived usability and perceived warm glow. Perceived usability was evaluated using the three dimensions outlined in ISO 9241-11: effectiveness, efficiency, and satisfaction. In the current context, perceived usability reflects users’ subjective evaluations of how well the system performs across these dimensions. In parallel, warm-glow motivation was measured to capture the emotional reward associated with engaging in behavior perceived as socially beneficial. This was divided into two subdimensions: intrinsic warm glow, which reflects internal feelings of moral satisfaction, and extrinsic warm glow, which involves a desire to be seen as prosocial by others. The five resulting constructs—perceived effectiveness (PEFF), perceived efficiency (PEFY), perceived satisfaction (PSTF), perceived extrinsic warm glow (PEWG), and perceived intrinsic warm glow (PIWG)—form the core of the measurement model. The items used to assess each construct, along with their sources, are listed in Table~\ref{tab:items}.

\begin{table*}[ht]
\centering
\caption{List of survey questions.}
\label{tab:items}
\begin{tabular}{p{2.2cm}p{1.4cm}p{8.0cm}p{4.0cm}}
\toprule
Construct & Item & Question & Source \\
\midrule
PEFF & PEFF1 & Using this search engine would improve my performance in my daily activities. 
             & Taken from Saravanos et al.~\cite{saravanos2022measuring} who adapted from Venkatesh and Bala~\cite{venkatesh2008}. \\
     & PEFF2 & Using this search engine would increase my productivity. & \\
     & PEFF3 & Using this search engine would enhance my effectiveness in my daily activities. & \\
     & PEFF4 & I would find this search engine to be useful in my daily activities. & \\
\addlinespace
PEFY & PEFY1 & I believe my interaction with this search engine would be clear and understandable. 
             & Taken from Saravanos et al.~\cite{saravanos2022measuring} who adapted from Venkatesh and Bala~\cite{venkatesh2008}. \\
     & PEFY2 & Interacting with this search engine would not require a lot of my mental effort. & \\
     & PEFY3 & I believe I would find this search engine to be easy to use. & \\
     & PEFY4 & I believe I would find it easy to get this search engine to do what I want it to do. & \\
\addlinespace
PSTF & PSTF1 & I would find using this search engine to be enjoyable. 
             & Taken from Saravanos et al.~\cite{saravanos2022measuring} who adapted from Venkatesh and Bala~\cite{venkatesh2008}. \\
     & PSTF2 & I would find the actual process of using this search engine to be pleasant. & \\
     & PSTF3 & I would have fun using this search engine. & \\
\addlinespace
PEWG & PEWG1 & Using this search engine fits the impression that I want to give to others that I am a good person whose actions have a positive impact on society. 
             & Saravanos et al.~\cite{saravanos2022measuring}. \\
     & PEWG2 & I am eager for my friends/acquaintances to learn about my use of this search engine and how I have a positive impact on society through its use. & \\
     & PEWG3 & I can express a more distinctive personal image through the use of this search engine by demonstrating to others that I am a good person who positively impacts society. & \\
\addlinespace
PIWG & PIWG1 & If I use this search engine, I will feel good because I do not only spend money for myself but also for other people. 
             & Saravanos et al.~\cite{saravanos2022measuring}. \\
     & PIWG2 & I feel comfortable if I donate for a good cause by using this search engine. & \\
     & PIWG3 & I am pleased that I do not only get a service by using this search engine, but that I also do a good deed at the same time. & \\
\bottomrule
\end{tabular}
\end{table*}

Each construct was measured using a multi-item Likert-type scale, with responses captured on a 7-point scale ranging from 1 (“Strongly disagree”) to 7 (“Strongly agree”). Composite variables were created by averaging the items within each construct. The four PEFF items, PEFF1–PEFF4, were adapted from Venkatesh and Bala~\cite{venkatesh2008} and used to measure users’ beliefs about how well the system supports task accomplishment. PEFF1 evaluates perceived performance improvement in daily activities, PEFF2 focuses on increased productivity, PEFF3 addresses perceived task effectiveness, and PEFF4 captures general system usefulness. The internal consistency of the PEFF scale was excellent, with a Cronbach’s $\alpha$ of 0.926.

The four perceived efficiency items, PEFY1–PEFY4, were also adapted from Venkatesh and Bala~\cite{venkatesh2008} and reflect users’ perceptions of the system’s ease of use and cognitive demand. PEFY1 assesses clarity and understandability of interactions, PEFY2 measures the perceived mental effort required, PEFY3 addresses general ease of use, and PEFY4 evaluates the user’s ability to get the system to perform as intended. The reliability of the PEFY scale was good ($\alpha = 0.844$).

Perceived satisfaction was measured by PSTF1–PSTF3, adapted from the construct of perceived enjoyment~\cite{venkatesh2008}. These items assess whether the system is experienced as pleasant or enjoyable, regardless of functional performance. PSTF1 evaluates overall enjoyment of use, PSTF2 focuses on the pleasantness of the interaction process, and PSTF3 captures the element of fun associated with use. The scale demonstrated excellent internal consistency ($\alpha = 0.950$).

PEWG also included three items, PEWG1–PEWG3, similarly adapted from Saravanos et al.~\cite{saravanos2022measuring}. These items measure the extent to which users perceive their behavior as socially visible and morally commendable. PEWG1 evaluates whether system use aligns with the user’s desired social image, PEWG2 assesses eagerness to share this behavior with others, and PEWG3 captures how system use contributes to a distinct, prosocial personal image. Internal consistency for this construct was also excellent ($\alpha = 0.927$).

The three PIWG items, PIWG1–PIWG3, were also adapted from Saravanos et al.~\cite{saravanos2022measuring} and capture the internal emotional satisfaction derived from contributing to a good cause through system use. PIWG1 assesses positive feelings from helping others, PIWG2 evaluates comfort with donating via system use, and PIWG3 reflects satisfaction with simultaneously receiving a service and doing good. This construct demonstrated excellent internal consistency ($\alpha = 0.921$).

All five constructs were analyzed by averaging responses to their corresponding items, forming composite scores that were used in subsequent statistical testing. Scale reliability statistics are summarized in Table~\ref{tab:alpha}. These results exceed commonly accepted thresholds, supporting the internal coherence of the scales and justifying their use in further analysis.

\begin{table*}[ht]
\centering
\caption{Internal consistency of construct measures.}
\label{tab:alpha}
\begin{tabular}{p{4cm}p{4cm}p{2cm}p{3cm}}
\toprule
Construct & Items Used & Cronbach’s $\alpha$ & Interpretation \\
\midrule
Perceived effectiveness & PEFF1, PEFF2, PEFF3, PEFF4 & 0.926 & Excellent \\
Perceived efficiency & PEFY1, PEFY2, PEFY3, PEFY4 & 0.844 & Good \\
Perceived satisfaction & PSTF1, PSTF2, PSTF3 & 0.950 & Excellent \\
Perceived extrinsic warm glow & PEWG1, PEWG2, PEWG3 & 0.927 & Excellent \\
Perceived intrinsic warm glow & PIWG1, PIWG2, PIWG3 & 0.921 & Excellent \\
\bottomrule
\end{tabular}
\end{table*}

\subsection{Participants and Procedure}

We collected data via an online questionnaire hosted on a secure survey platform. The 164 participants first completed demographic and background questions, followed by the questions included to capture user perceptions. We attempted to randomize questions to reduce potential order effects relying on the survey platform’s functionality. Responses were monitored for completeness and consistency, with incomplete responses or those exhibiting patterns of inattentiveness excluded from the final dataset. Gender distribution was nearly equal, with 51.8\% identifying as female and 48.2\% as male. Participants aged 31–55 comprised 78.0\% of the sample, while those aged 26–30 and 18–25 accounted for 10.4\% and 3.0\%, respectively; respondents aged 56 or older made up 8.5\%. Reported household income spanned five categories from \$\textless 19{,}999 to $\geq 150{,}000$, with 34\% of participants in the \$50{,}000–\$79{,}999 range. Educational attainment was distributed as follows: 37.8\% held a bachelor’s degree, 17.1\% had attended college without earning a degree, 15.9\% had a high-school diploma or equivalent, 14.6\% held an associate degree, 14.0\% held a graduate degree, and 0.6\% had less than a high school education. The range of demographic characteristics suggests that the findings are relevant to adults with varied backgrounds.

\section{Results}
\label{sec:results}

To examine how warm glow influences perceived usability, a structured series of statistical analyses was conducted using Python and its associated libraries, including the \texttt{pandas}, \texttt{statsmodels}, \texttt{pingouin}, and \texttt{scipy} packages. This section begins with a MANOVA to assess the joint effects of PIWG and PEWG across all usability dimensions. It then reports univariate ANOVAs to isolate effects on individual outcomes. To ensure robustness, both permutation-based significance testing and bootstrapped confidence intervals were calculated, offering nonparametric validation of the findings. We outline the results of each analysis in the following subsections.

\subsection{Multivariate Analysis of Perceived Intrinsic and Extrinsic Warm Glow}

A MANOVA was conducted to examine how PEWG, PIWG, and their interaction influenced participants’ combined ratings of perceived effectiveness, perceived efficiency, and perceived satisfaction. Classical (parametric) MANOVA results indicated that PIWG had a significant multivariate effect across the three outcomes ($F(3,158) = 8.076$, $p < 0.001$, Pillai’s trace $= 0.133$). PEWG also exhibited a significant multivariate effect ($F(3,158) = 5.677$, $p = 0.001$, Pillai’s trace $= 0.097$). Moreover, the interaction between PIWG and PEWG was significant ($F(3,158) = 3.447$, $p = 0.018$, Pillai’s trace $= 0.061$), meaning that participants who highly rated both strong internal satisfaction and strong external recognition tended to give higher overall ratings on perceived effectiveness, efficiency, and satisfaction than would be expected from either factor alone.

To assess robustness, permutation and bootstrapping procedures were undertaken. A permutation-based MANOVA with 10{,}000 resamples revealed that PIWG remained highly significant (observed $F = 8.076$, permutation $p = 0.000$), as did PEWG (observed $F = 5.677$, permutation $p = 0.000$) and their interaction (observed $F = 3.447$, permutation $p = 0.017$). These permutation results support the conclusion that the multivariate effects are unlikely to arise by chance given the data’s distribution.

Bootstrap confidence intervals for the $F$-statistics were generated from 10{,}000 resamples to further evaluate stability. The 95\% confidence interval for Pillai’s trace associated with PIWG ranged from 0.051 to 0.287, while the interval for its $F$-statistic spanned 2.812 to 21.186. For PEWG, the 95\% confidence interval for Pillai’s trace was 0.028 to 0.252, and the $F$-statistic fell between 1.494 and 17.720. Finally, the interaction between PIWG and PEWG yielded a Pillai’s trace interval of 0.0129 to 0.1980 and an $F$-statistic interval of 0.690 to 13.002. Taken together, these bootstrap intervals indicate that the multivariate effects of PIWG, PEWG, and their interaction are consistently positive and of non-trivial magnitude, rather than artifacts of sampling variability. These results are summarized in Table~\ref{tab:manova}.

\begin{table}[ht]
\centering
\caption{MANOVA results.}
\label{tab:manova}
\begin{tabular}{lcccc}
\toprule
Effect & Pillai’s trace & $F$-value & df & $p$-value \\
\midrule
PIWG & 0.133 & 8.076 & 3, 158 & $< 0.001$ \\
PEWG & 0.097 & 5.677 & 3, 158 & 0.001 \\
PIWG $\times$ PEWG & 0.061 & 3.447 & 3, 158 & 0.018 \\
\bottomrule
\end{tabular}
\end{table}

\subsection{Univariate Analyses for Each Outcome}

Next, each outcome variable (perceived effectiveness, perceived efficiency, and perceived satisfaction) was examined using a series of classical analysis of variance (ANOVA) alongside permutation and bootstrap methods to assess the robustness and consistency of the observed effects (Table~\ref{tab:anova_effectiveness}--\ref{tab:anova_satisfaction}). For perceived effectiveness, the ANOVA indicated significant main effects of PIWG and PEWG, as well as a significant interaction between them (PIWG: $F = 6.036$, $p = 0.015$; PEWG: $F = 6.853$, $p = 0.010$; interaction: $F = 7.507$, $p = 0.007$). Permutation tests yielded corresponding permutation $p$-values of 0.015 for PIWG, 0.010 for PEWG, and 0.008 for their interaction, confirming that these effects were unlikely to have arisen by chance given the data’s distribution. Bootstrapped confidence intervals for the $F$-statistics further supported these findings, with 95\% confidence intervals (CIs) for PIWG, PEWG, and their interaction of [0.050, 26.085], [0.663, 21.940], and [0.093, 29.126], respectively. Across 10,000 resampled datasets, the confidence intervals for the F‑statistics were consistently centered on moderate to large values, reinforcing the conclusion that intrinsic warm glow, extrinsic warm glow, and their interaction have non‑negligible effects on perceived effectiveness.

\begin{table*}[ht]
\centering
\caption{Univariate ANOVA for perceived effectiveness.}
\label{tab:anova_effectiveness}
\begin{tabular}{lcccc}
\toprule
Predictor & $F$-value & $p$-value & Permutation $p$-value & Bootstrapped 95\% CI for $F$ \\
\midrule
PIWG & 6.036 & 0.015 & 0.015 & [0.050, 26.085] \\
PEWG & 6.853 & 0.010 & 0.010 & [0.663, 21.940] \\
PIWG $\times$ PEWG & 7.507 & 0.007 & 0.008 & [0.093, 29.126] \\
\bottomrule
\end{tabular}
\end{table*}

For perceived efficiency, the classical ANOVA showed that only PIWG was a significant predictor for efficiency ($F = 5.269$, $p = 0.023$), whereas PEWG and the interaction term were not significant (Table~\ref{tab:anova_efficiency}). Permutation testing echoed this result, with PIWG’s observed $F$-statistic yielding a permutation $p$-value of 0.0251, while those for PEWG and the interaction were 0.5776 and 0.2014, respectively. Bootstrapped 95\% confidence intervals for the $F$-statistic were [0.123, 19.398] for PIWG, [0.001, 5.819] for PEWG, and [0.005, 11.853] for their interaction. Because PIWG’s interval was clearly positive, bootstrap resampling confirms that the positive influence of intrinsic warm glow on perceived efficiency is a stable feature of the data. In contrast, the intervals for PEWG and the interaction extended down to very small F-values, consistent with their nonsignificant results under the classical and permutation tests.

\begin{table*}[ht]
\centering
\caption{Univariate ANOVA for perceived efficiency.}
\label{tab:anova_efficiency}
\begin{tabular}{lcccc}
\toprule
Predictor & $F$-value & $p$-value & Permutation $p$-value & Bootstrapped 95\% CI for $F$ \\
\midrule
PIWG & 5.269 & 0.023 & 0.025 & [0.123, 19.398] \\
PEWG & 0.314 & 0.578 & 0.578 & [0.001, 5.819] \\
PIWG $\times$ PEWG & 1.632 & 0.201 & 0.201 & [0.005, 11.853] \\
\bottomrule
\end{tabular}
\end{table*}

For perceived satisfaction, classical ANOVA revealed significant effects of both PIWG ($F = 35.153$, $p < 0.001$) and PEWG ($F = 15.013$, $p < 0.001$), with their interaction remaining nonsignificant (Table~\ref{tab:anova_satisfaction}). Permutation tests corroborated these findings: the observed $F$-statistics for PIWG, PEWG, and their interaction produced permutation $p$-values of $< 0.000$, 0.000, and 0.195, respectively. Bootstrapped 95\% confidence intervals for the $F$-statistics were [11.648, 74.345] for PIWG, [2.594, 38.782] for PEWG, and [0.005, 12.746] for their interaction. Since the intervals for both PIWG and PEWG were clearly positive, the bootstrap analysis confirmed that each form of warm glow exerts a reliable, positive influence on satisfaction. For the interaction, the interval extended down to very small F-values, consistent with its lack of statistical significance.

\begin{table*}[ht]
\centering
\caption{Univariate ANOVA for perceived satisfaction.}
\label{tab:anova_satisfaction}
\begin{tabular}{lcccc}
\toprule
Predictor & $F$-value & $p$-value & Permutation $p$-value & Bootstrapped 95\% CI for $F$ \\
\midrule
PIWG & 35.153 & $< 0.001$ & $< 0.000$ & [11.648, 74.345] \\
PEWG & 15.013 & $< 0.001$ & 0.000 & [2.594, 38.782] \\
PIWG $\times$ PEWG & 1.666 & 0.195 & 0.195 & [0.005, 12.746] \\
\bottomrule
\end{tabular}
\end{table*}

Taken together, the classical and permutation analyses consistently demonstrate that intrinsic warm glow significantly affects all three outcomes, while extrinsic warm glow significantly influences perceived effectiveness and perceived satisfaction but not perceived efficiency. The interaction between intrinsic and extrinsic warm glow is particularly important for perceived effectiveness, where their combination yields the largest gains relative to either factor alone. Bootstrapped confidence intervals for the F-statistics were strictly positive for the significant effects and extended towards very small values for the nonsignificant ones, reinforcing the classical and permutation-based findings and supporting the robustness of the observed relationships.

\section{Discussion and Conclusions}
\label{sec:discussion}

This study contributes to a growing body of research on human–computer interaction by demonstrating that warm glow can act as a factor in shaping usability perceptions. Our findings demonstrate that users who experience warm glow tend to report higher satisfaction and more favorable efficiency and effectiveness judgments in usability appraisals, thereby bridging the gap between classical usability research and user warm-glow feelings. In this study, we examined how the two dimensions of warm glow, extrinsic warm glow arising from anticipated social recognition and intrinsic warm glow arising from personal moral satisfaction, can influence how an individual perceives the usability of a technology.

The work expands on the traditional approach to measuring usability, which focuses on the functional aspect of a system. Regarding the effect of intrinsic warm glow on perceived usability, specifically effectiveness, efficiency, and satisfaction (RQ1), the results show that intrinsic warm glow enhances all three dimensions. This finding supports the work of Saravanos et al.~\cite{saravanos2026utaut2wg}, who found a similar impact on all three usability dimensions. Therefore, a technology that evokes intrinsic warm glow is perceived by users as more useful, easier to use, and resulting in more satisfaction from its use.

In contrast, extrinsic warm glow was found to have a selective effect on perceived usability across the same dimensions (RQ2). It significantly influenced perceived effectiveness and perceived satisfaction but not perceived efficiency. In other words, when people believed that using the technology would improve their social image (e.g., being seen as someone who does good), they tended to view it as more valuable and enjoyable but not necessarily easier to use. These results echo previous investigations of the warm-glow phenomenon. Saravanos et al.~\cite{saravanos2022tam3wg} found that users feeling extrinsic warm glow have a more positive view of the perceived usefulness of a technology. Saravanos et al.~\cite{saravanos2026utaut2wg} also found that extrinsic warm glow influences the perceived usefulness of the technology as well as the perceived hedonic motivation derived from its use. Our results also complement existing work in human–computer interaction, which has shown that usability can be altered by feelings, such as those of aesthetics~\cite{kurosu1995,norman2002,tractinsky1997}. For example, looking at a technology whose design evokes a positive feeling can influence usability.

With respect to how intrinsic and extrinsic warm glow interact in shaping perceived usability (RQ3), the findings revealed a statistically significant interaction effect between intrinsic and extrinsic warm glow on perceived effectiveness. Specifically, when both forms of warm glow were high, participants rated the system as more effective than would be expected based on either form alone. This suggests a synergistic effect, where moral alignment and social validation reinforce one another to enhance perceptions of effectiveness. In contrast, this interaction effect was not present for perceived efficiency or perceived satisfaction, indicating that those dimensions may be more independently influenced by each form of warm glow.

\subsection{Theoretical Implications}

The findings of this work support and expand the current usability literature, which focuses on design (Norman~\cite{norman2002}, Kurosu and Kashimura~\cite{kurosu1995}, and Tractinsky~\cite{tractinsky1997}) and values (see Friedman et al.~\cite{friedman2013}). First, it demonstrates that warm glow can influence perceived usability. Moreover, it identifies which types of warm glow influence which types of usability, supporting previous work~\cite{saravanos2026utaut2wg,saravanos2022tam3wg}. Evoking intrinsic warm glow influences all three usability dimensions (i.e., effectiveness, efficiency, and satisfaction), whereas evoking extrinsic warm glow influences only perceived effectiveness and perceived satisfaction and not perceived efficiency. Moreover, we demonstrate that one can craft experiences that make users feel a “warm glow”; we show it is possible to engineer positive affective states that improve the user’s interaction with a system. This extends prior work on emotional design, similar to how the aesthetic-usability effect shows that users perceive attractive interfaces as more usable~(i.e., ~\cite{norman2002, kurosu1995, tractinsky1997, friedman2013}).

\subsection{Practical Implications}

The study underscores that warm glow plays a measurable role in shaping what are typically considered rational judgments about technology. Practically, our findings have clear design implications. To improve the user experience, designers can actively create opportunities for users to experience warm glow. In actionable terms, this means designs that allow users to see their positive impact on others or align with a prosocial cause as a seamless part of the user journey, for example, by embedding features or interaction flows. When users know that using a system contributes to something they personally value, such as a charitable cause, community well-being, or another person’s benefit, they are likely to experience warm glow, which may translate into a more forgiving and engaged mindset toward the product.

Designers of technology should therefore regard warm-glow evocation as an explicit usability goal. Just as they strive to minimize cognitive load or streamline task flows, they can also consider how we might elicit warm-glow feelings in users. Simple design elements, such as feedback highlighting a user’s contribution to a shared good or narratives that frame actions in terms of helping others, can trigger warm glow when done authentically. Such elements should be woven into the experience in a way that feels natural and integral to the system’s purpose, thereby reinforcing the user’s values through their normal use of the system. Thus, by embedding characteristics that will evoke warm-glow feelings among users into the design, developers and organizations may be able to enhance perceived usability, not by changing the functionality itself but by influencing how users emotionally and socially relate to the technology.

Designers seeking to foster intrinsic warm glow should consider embedding features that visibly and meaningfully align with users’ values. To effectively activate extrinsic warm glow, designers could incorporate social visibility and validation into their interfaces, such as enabling easy sharing of positive user impacts on social media, providing digital badges, or using leaderboards to highlight prosocial achievements publicly. For example, if a user perceives that a technology is too difficult to use under normal conditions, warm-glow motivations may encourage continued use. This could occur when users know their use contributes to their favorite charity (intrinsic motivation) or when they expect peer recognition for using the technology (i.e., extrinsic motivation). While motivations that evoke warm glow may differ between individuals, certain patterns can often be generalized within particular groups of users.

\subsection{Limitations and Next Steps}

While this study offers novel insights into expanding usability to address user feelings, it also has several limitations that suggest opportunities for future research. First, the sample was collected exclusively from participants in the United States. Prior research suggests that usability may be influenced by culture (e.g., Kurosu and Kashimura~\cite{kurosu1995}, Norman~\cite{norman2002}, and Tractinsky~\cite{tractinsky1997}). As such, the generalizability of the findings may be limited to an American cultural context. Moreover, even within the United States there may be differences. Therefore, future studies could investigate differences at both individual and societal levels. One approach could be to examine what evokes warm glow in different groups, cities, states. Another approach could be to incorporate participants from diverse cultural backgrounds to explore how the exchange between usability and warm-glow perceptions is similar or varies among societies.

Second, the study focused on a single technological context, specifically an internet search engine that is free, easily recognizable, accessible, and easy to use. While this choice offered experimental control, warm-glow effects may differ for technologies with different characteristics. Replicating this study with diverse technologies would help clarify whether domain-specific factors mediate the impact of warm glow.

Third, this study offers only a snapshot in time. Warm-glow responses may evolve with repeated exposure, user goals, and social norms. Longitudinal studies are needed to assess whether motivational effects persist, fade, or transform as users become familiar with a system or as the novelty of prosocial features fades. This is in line with Bargas-Avila and Hornbæk~\cite{bargasavila2011}, who note that emotional outcomes from using a technology may change over time.

Despite these limitations, this study makes an important contribution to the field of human–computer interaction by demonstrating that both intrinsic and extrinsic warm-glow meaningfully shape how users perceive the usability of technology. By moving beyond traditional functional definitions of usability and showing that affective, value-driven experiences can enhance users’ evaluations of effectiveness, efficiency, and satisfaction, our findings open new avenues for research and design. Ultimately, this work underscores the importance of designing technology that resonates with users’ values and emotions, not just their functional needs. This approach offers a pathway to more engaging, meaningful, and widely adopted technologies.

\end{document}